\documentstyle[prl,aps,preprint,epsf,floats]{revtex}
\begin{document}
\preprint{MIT-CTP-2564}

\draft
\def\footnoterule{\kern-3pt \hrule width\hsize \kern3pt}
\tighten
\title{Feynman diagrams with the effective action}

\author{M.J. de la Plata${}^1$ and L.L. Salcedo${}^{1,2}$}

\address{
{~} \\
${}^1$Departamento de F\'{\i}sica Moderna \\
Universidad de Granada \\
E-18071 Granada, Spain \\
{~} \\
${}^2$Center for Theoretical Physics \\
Laboratory for Nuclear Science \\
and Department of Physics \\
Massachusetts Institute of Technology \\
Cambridge, Massachusetts 02139, U.S.A. \\
{~}}

\date{Received \today}
\maketitle

\thispagestyle{empty}

\begin{abstract}

A derivation is given of the Feynman rules to be used in the
perturbative computation of the Green's functions of a generic quantum
many-body theory when the action which is being perturbed is not
necessarily quadratic. Some applications are discussed.

\end{abstract}



\section{Introduction}
The Feynman diagrammatic technique has proven quite useful in order to
perform and organize the perturbative solution of quantum many-body
theories. The main idea is the computation of the Green's or
correlation functions by splitting the action $S$ into a quadratic or
free part $S_Q$ plus a remainder or interacting part $S_I$ which is
then treated as a perturbation. From the beginning this technique has
been extended to derive exact relations, such as the
Schwinger-Dyson~\cite{Dy49,Sc51,It80} equations, or to make
resummation of diagrams as that implied in the effective action
approach~\cite{Il75,Ne87} and its generalizations~\cite{Co74}.

Consider now a generalization of the above problem, namely, to solve
(i.e., to find the Green's functions of) a theory with action given by
$S+\delta S$ perturbatively in $\delta S$ but where the
``unperturbed'' action $S$ (assumed to be solved) is not necessarily
quadratic in the fields. The usual answer to this problem is to write
the action as a quadratic part $S_Q$ plus a perturbation $S_I+\delta
S$ and then to apply the standard Feynman diagrammatic technique. This
approach is, of course, correct but it does not exploit the fact that
the unperturbed theory $S$ is solved, i.e., its Green's functions are
known. For instance, the computation of each given order in $\delta S$
requires an infinite number of diagrams to all orders in $S_I$. We
will refer to this as the {\em standard expansion}. In this paper it
is shown how to systematically obtain the Green's functions of the
full theory, $S+\delta S$, in terms of those of the unperturbed one,
$S$, plus the vertices provided by the perturbation, $\delta
S$. Unlike the standard expansion, in powers of $S_I+\delta S$, the
expansion considered here is a strict perturbation in $\delta S$ and
constitutes the natural extension of the Feynman diagrammatic
technique to unperturbed actions which are not necessarily
quadratic. We shall comment below on the applications of such an
approach.

\section{Many-body theory background}
\subsection{Feynman diagrams and standard Feynman rules}
In order to state our general result let us recall some well known
ingredients of quantum many-body theory (see e.g.~\cite{Ne87}), and in
passing, introduce some notation and give some needed definitions.
Consider an arbitrary quantum many-body system described by variables
or {\em fields} $\phi^i$, that for simplicity in the presentation will
be taken as bosonic. As will be clear below, everything can be
generalized to include fermions. Without loss of generality we can use
a single discrete index $i$ to represent all the needed labels (DeWitt
notation). For example, for a relativistic quantum field theory, $i$
would contain space-time, Lorentz and Dirac indices, flavor, kind of
particle and so on. Within a functional integral formulation of the
many-body problem, the expectation values of observables, such as
$A[\phi]$, take the following form:
\begin{equation}
\langle A[\phi] \rangle = \frac{\int\exp\left(S[\phi]\right)A[\phi]\,d\phi}
{\int\exp\left(S[\phi]\right)\,d\phi}\,.
\label{eq:1}
\end{equation}
Here the function $S[\phi]$ will be called the {\em action} of the
system and is a functional in general. Note that in some cases
$\langle A[\phi]\rangle$ represents the time ordered vacuum
expectation values, in other the canonical ensemble averages, etc, and
also the quantity $S[\phi]$ may correspond to different objects in
each particular application. In any case, all (bosonic) quantum
many-body systems can be brought to this form and only
eq.~(\ref{eq:1}) is needed to apply the Feynman diagrammatic
technique. As already noted, this technique corresponds to write the
action in the form $S[\phi]=S_Q[\phi]+S_I[\phi]$:
\begin{equation}
S_Q[\phi]=\frac{1}{2}m_{ij}\phi^i\phi^j\,,\qquad 
S_I[\phi]=\sum_{n\geq 0}\frac{1}{n!}g_{i_1\dots i_n}
\phi^{i_1}\cdots\phi^{i_n} \,,
\end{equation}
where we have assumed that the action is an analytical function of the
fields at $\phi^i=0$. Also, a repeated indices convention will be used
throughout. The quantities $g_{i_1\dots i_n}$ are the {\em coupling
constants}. The matrix $m_{ij}$ is non singular and otherwise
arbitrary, whereas the combination $m_{ij}+g_{ij}$ is completely
determined by the action. The {\em free propagator}, $s^{ij}$, is
defined as the inverse matrix of $-m_{ij}$. The signs in the
definitions of $S[\phi]$ and $s^{ij}$ have been chosen so that there
are no minus signs in the Feynman rules below. The $n$-point {\em
Green's function} is defined as
\begin{equation}
G^{i_1\cdots i_n}= \langle\phi^{i_1}\cdots\phi^{i_n}\rangle\,, \quad n\geq 0\,.
\end{equation}
Let us note that under a non singular linear transformation of the
fields, and choosing the action to be a scalar, the coupling constants
transform as completely symmetric covariant tensors and the propagator
and the Green's functions transform as completely symmetric
contravariant tensors. The tensorial transformation of the Green's
functions follows from eq.~(\ref{eq:1}), since the constant Jacobian
of the transformation cancels among numerator and denominator.

\begin{figure}
\begin{center}
\vspace{-0.cm}                                                                
\leavevmode
\epsfysize = 2.0cm
\makebox[0cm]{\epsfbox{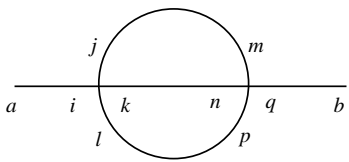}}
\end{center}
\caption{Feynman graph corresponding to the monomial in
eq.~(\ref{eq:example}).} 
\label{f:1}
\end{figure}

Perturbation theory consists of computing the Green's functions as a
Taylor expansion in the coupling constants. We remark that the
corresponding series is often asymptotic, however, the perturbative
expansion is always well defined. By inspection, and recalling the
tensorial transformation properties noted above, it follows that the
result of the perturbative calculation of $G^{i_1\cdots i_n}$ is a sum
of monomials, each of which is a contravariant symmetric tensor
constructed with a number of coupling constants and propagators, with
all indices contracted except $(i_1\cdots i_n)$ times a purely
constant factor. For instance,
\begin{equation}
G^{ab}= \cdots + \frac{1}{3!}s^{ai}g_{ijk\ell}s^{jm}s^{kn}s^{\ell
p}g_{mnpq}s^{qb} +\cdots \,.
\label{eq:example}
\end{equation}
Each monomial can be represented by a {\em Feynman diagram} or graph:
each $k$-point coupling constant is represented by a vertex with $k$
prongs, each propagator is represented by an unoriented line with two
ends. The dummy indices correspond to ends attached to vertices and
are called {\em internal}, the free indices correspond to unattached
or external ends and are the {\em legs} of the diagram. The lines
connecting two vertices are called {\em internal}, the others are {\em
external}. By construction, all prongs of every vertex must be
saturated with lines. The diagram corresponding to the monomial in
eq.~(\ref{eq:example}) is shown in figure~\ref{f:1}.

\begin{figure}[htb]
\centerline{\mbox{\epsfysize=4.0cm\epsffile{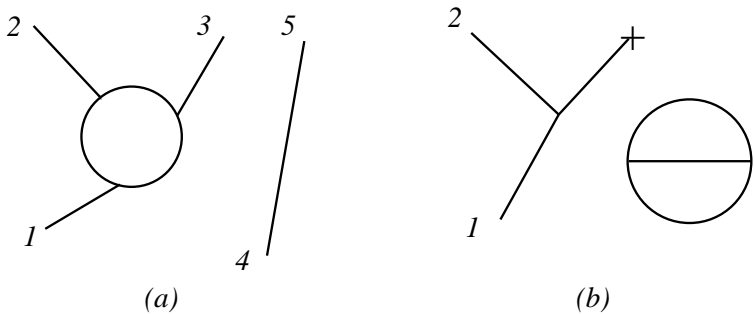}}}
\vspace{6pt}
\caption{$(a)$ A linked disconnected graph. $(b)$ A unlinked
graph. The cross represents a 1-point vertex.}
\label{f:2}
\end{figure}

A graph is {\em connected} if it is connected in the topological
sense. A graph is {\em linked} if every part of it is connected to at
least one of the legs (i.e., there are no disconnected $0$-legs
subgraphs). All connected graphs are linked. For instance, the graph
in figure~\ref{f:1} is connected, that in figure~\ref{f:2}$a$ is
disconnected but linked and that in figure~\ref{f:2}$b$ is
unlinked. To determine completely the value of the graph, it only
remains to know the weighting factor in front of the monomial. As
shown in many a textbook~\cite{Ne87}, the factor is zero if the
diagram is not linked. That is, unlinked graphs are not to be included
since they cancel due to the denominator in eq.~(\ref{eq:1}); a result
known as Goldstone theorem. For linked graphs, the factor is given by
the inverse of the {\em symmetry factor} of the diagram which is
defined as the order of the symmetry group of the graph. More
explicitly, it is the number of topologically equivalent ways of
labeling the graph. For this counting all legs are distinguishable
(due to their external labels) and recall that the lines are
unoriented. Dividing by the symmetry factor ensures that each distinct
contributions is counted once and only once. For instance, in
figure~\ref{f:1} there are three equivalent lines, hence the factor
$1/3!$ in the monomial of eq.~(\ref{eq:example}).

Thus, we arrive to the following {\em Feynman rules} to compute
$G^{i_1\cdots i_n}$ in perturbation theory:
\begin{enumerate}
\item Consider each $n$-point linked graph. Label the legs with
$(i_1,\dots,i_n)$, and label all internal ends as well.
\item Put a factor $g_{j_1\dots j_k}$ for each $k$-point vertex, and a
factor $s^{ij}$ for each line. Sum over all internal indices and
divide the result by the symmetry factor of the graph.
\item Add up the value of all topologically distinct such graphs.
\end{enumerate}
We shall refer to the above as the Feynman rules of the theory
``$S_Q+S_I$''. There are several relevant remarks to be made: If
$S[\phi]$ is a polynomial of degree $N$, only diagrams with at most
$N$-point vertices have to be retained. The choice $g_{ij}=0$ reduces
the number of diagrams. The 0-point vertex does not appear in any
linked graph. Such term corresponds to an additive constant in the
action and cancels in all expectation values. On the other hand, the
only linked graph contributing to the 0-point Green's function is a
diagram with no elements, which naturally takes the value 1.

Let us define the {\em connected Green's functions}, $G_c^{i_1\cdots
i_n}$, as those associated to connected graphs (although they can be
given a non perturbative definition as well). From the Feynman rules
above, it follows that linked disconnected diagrams factorize into its
connected components, thus the Green's functions can be expressed in
terms of the connected ones. For instance
\begin{eqnarray}
G^i &=& G_c^i \,, \nonumber  \\
G^{ij} &=& G_c^{ij} + G_c^iG_c^j \,,  \\
G^{ijk} &=& G_c^{ijk} +
G_c^iG_c^{jk} + G_c^jG_c^{ik} + G_c^kG_c^{ij} + G_c^iG_c^jG_c^k \,. 
\nonumber 
\end{eqnarray}

It will also be convenient to introduce the {\em generating function}
of the Green's functions, namely,
\begin{equation}
Z[J] = \int\exp\left(S[\phi]+J\phi\right)\,d\phi \,,
\label{eq:Z}
\end{equation}
where $J\phi$ stands for $J_i\phi^i$ and $J_i$ is called the {\em
external current}. By construction,
\begin{equation}
\frac{Z[J]}{Z[0]} = \langle\exp\left(J\phi\right)\rangle
=\sum_{n\geq 0}\frac{1}{n!}G^{i_1\cdots i_n}J_{i_1}\cdots J_{i_n}\,,
\end{equation}
hence the name generating function. The quantity $Z[0]$ is known as
{\em partition function}. Using the replica method~\cite{Ne87}, it can
be shown that $W[J]=\log\left(Z[J]\right)$ is the generator of the
connected Green's functions. It is also shown that $W[0]$ can be
computed, within perturbation theory, by applying essentially the same
Feynman rules given above as the sum of connected diagrams without
legs and the proviso of assigning a value $-\frac{1}{2}{\rm
tr}\,\log(-m/2\pi)$ to the diagram consisting of a single closed
line. The partition function is obtained if non connected diagrams are
included as well. In this case, it should be noted that the
factorization property holds only up to possible symmetry factors.

\subsection{The effective action}
\begin{figure}[htb]
\centerline{\mbox{\epsfysize=3.5cm\epsffile{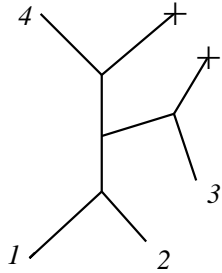}}}
\caption{A tree graph.}
\label{f:3}
\end{figure}

To proceed, let us introduce the {\em effective action}, which will be
denoted $\Gamma[\phi]$. It can be defined as the Legendre transform of
the connected generating function. For definiteness we put this in the
form
\begin{equation}
\Gamma[\phi] = \min_J\left(W[J]-J\phi\right)\,,
\end{equation}
although in general $S[\phi]$, $W[J]$, as well as the fields, etc, may
be complex and only the extremal (rather than minimum) property is
relevant. For perturbation theory, the key feature of the effective
action is as follows. Recall that a connected graph has $n$ {\em
loops} if it is possible to remove at most $n$ internal lines so that
it remains connected. For an arbitrary graph, the number of loops is
defined as the sum over its connected components. {\em Tree} graphs
are those with no loops. For instance the diagram in
figure~\ref{f:1} has two loops whereas that in figure~\ref{f:3} is
a tree graph. Then, the effective action coincides with the equivalent
action that at tree level would reproduce the Green's functions of
$S[\phi]$. To be more explicit, let us make an arbitrary splitting of
$\Gamma[\phi]$ into a (non singular) quadratic part $\Gamma_Q[\phi]$
plus a remainder, $\Gamma_I[\phi]$,
\begin{equation}
\Gamma_Q[\phi]=\frac{1}{2}\overline{m}_{ij}\phi^i\phi^j\,, \qquad
\Gamma_I[\phi]=\sum_{n\ge 0}\frac{1}{n!}
\overline{g}_{i_1\dots i_n}\phi^{i_1}\cdots\phi^{i_n}\,,
\end{equation}
then the Green's functions of $S[\phi]$ are recovered by using the
Feynman rules associated to the theory ``$\Gamma_Q+\Gamma_I$'' but
adding the further prescription of including only tree level
graphs. The building blocks of these tree graphs are the {\em
effective line}, $\overline{s}^{ij}$, defined as the inverse matrix of
$-\overline{m}_{ij}$, and the {\em effective (or proper) vertices},
$\overline{g}_{i_1\dots i_n}$. This property of the effective action
will be proven below. Let us note that $\Gamma[\phi]$ is completely
determined by $S[\phi]$, and is independent of how $m_{ij}$ and
$\overline{m}_{ij}$ are chosen. In particular, the combination
$\overline{m}_{ij}+\overline{g}_{ij}$ in free of any choice.  Of
course, the connected Green's are likewise obtained at tree level from
the theory ``$\Gamma_Q+\Gamma_I$'', but including only connected
graphs.
For ulterior reference, let us define the {\em effective current} as
$\overline{g}_i$ and the {\em self-energy} as
\begin{equation}
\Sigma_{ij}= \overline{m}_{ij}+\overline{g}_{ij}-m_{ij}\,.
\end{equation}
Note that $\Sigma_{ij}$ depends not only on $S[\phi]$ but also on the
choice of $S_Q[\phi]$.

\begin{figure}[htb]
\centerline{\mbox{\epsfysize=3.5cm\epsffile{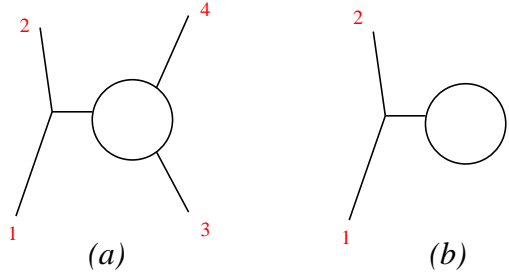}}}
\caption{$(a)$ A 1-particle reducible graph. $(b)$ A graph with a
tadpole subgraph.}
\label{f:4}
\end{figure}

A connected graph is {\em 1-particle irreducible} if it remains
connected after removing any internal line, and otherwise it is called
{\em 1-particle reducible}. In particular, all connected tree graphs
with more than one vertex are reducible. For instance the graph in
figure~\ref{f:1} is 1-particle irreducible whereas those in
figures~\ref{f:3} and ~\ref{f:4} are reducible. To {\em amputate} a
diagram (of the theory ``$S_Q+S_I$'') is to contract each leg with a
factor $-m_{ij}$. In the Feynman rules, this corresponds to not to
include the propagators of the external legs. Thus the amputated
diagrams are covariant tensors instead of contravariant. Then, it is
shown that the $n$-point effective vertices are given by the connected
1-particle irreducible amputated $n$-point diagrams of the theory
``$S_Q+S_I$''. (Unless $n=2$. In this case the sum of all such
diagrams with at least one vertex gives the self-energy.)

\begin{figure}[htb]
\centerline{\mbox{\epsfysize=4.5cm\epsffile{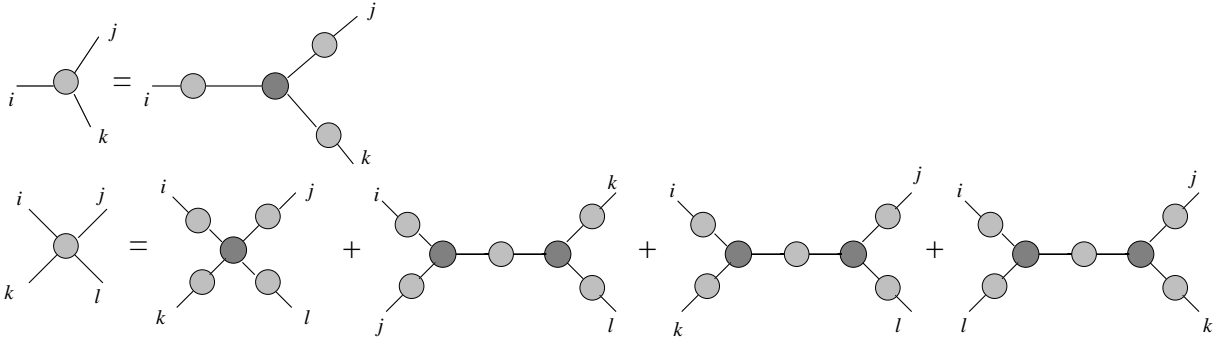}}}
\vspace{6pt}
\caption{Feynman diagrams for the 3- and 4-point connected Green's
functions in terms of the proper functions
(cf. eq.~(\ref{eq:connected})). The lighter blobs represent the
connected functions, the darker blobs represent the irreducible
functions.}
\label{f:5}
\end{figure}

A graph has {\em tadpoles} if it contains a subgraph from which stems
a single line. It follows that all graphs with 1-point vertices have
tadpoles. Obviously, when the single line of the tadpole is internal,
the graph is 1-particle reducible (cf. figure~\ref{f:4}$b$). An
important particular case is that of actions for which
$\langle\phi^i\rangle$ vanishes. This ensures that the effective
current vanishes, i.e. $\overline{g}_i=0$ and thus all tree graphs of
the theory ``$\Gamma_Q+\Gamma_I$'' are free of tadpoles (since tadpole
subgraphs without 1-point vertices require at least one loop). Given
any action, $\langle\phi^i\rangle=0$ can be achieved by a redefinition
of the field $\phi^i$ by a constant shift, or else by a readjustment
of the original current $g_i$, so this is usually a convenient
choice. A further simplification can be achieved if $\Gamma_Q[\phi]$
is chosen as the full quadratic part of the effective action, so that
$\overline{g}_{ij}$ vanishes. Under these two choices, each Green's
function requires only a finite number of tree graphs of the theory
``$\Gamma_Q+\Gamma_I$''. Also, $\overline{s}^{ij}$ coincides with the
full connected propagator, $G_c^{ij}$, since a single effective line
is the only possible diagram for it. Up to 4-point functions, it is
found
\begin{eqnarray}
G_c^i &=& 0 \,, \nonumber \\
G_c^{ij} &=& \overline{s}^{ij} \,, \label{eq:connected}
\\
G_c^{ijk} &=&
\overline{s}^{ia}\overline{s}^{jb}\overline{s}^{kc}\overline{g}_{abc}
\,,\nonumber \\
G_c^{ijk\ell} &=&
\overline{s}^{ia}\overline{s}^{jb}\overline{s}^{kc}\overline{s}^{\ell
d}\overline{g}_{abcd} \nonumber \\
& & +\overline{s}^{ia}\overline{s}^{jb}\overline{g}_{abc}
\overline{s}^{cd}\overline{g}_{def}\overline{s}^{ek}\overline{s}^{f\ell}
+\overline{s}^{ia}\overline{s}^{kb}\overline{g}_{abc}
\overline{s}^{cd}\overline{g}_{def}\overline{s}^{ej}\overline{s}^{f\ell} 
+\overline{s}^{ia}\overline{s}^{\ell b}\overline{g}_{abc}
\overline{s}^{cd}\overline{g}_{def}\overline{s}^{ek}\overline{s}^{fj}
\,.\nonumber
\end{eqnarray}
The corresponding diagrams are depicted in figure~\ref{f:5}. Previous
considerations imply that in the absence of tadpoles, $G_c^{ij}=
-((m+\Sigma)^{-1})^{ij}$.

\section{Perturbation theory on non quadratic actions}
\subsection{Statement of the problem and main result}
All the previous statements are well known in the literature. Consider
now the action $S[\phi]+\delta S[\phi]$, where
\begin{equation}
\delta S[\phi]=\sum_{n\ge 0}\frac{1}{n!} \delta g_{i_1\dots i_n}
\phi^{i_1}\cdots\phi^{i_n}\,,
\end{equation}
defines the {\em perturbative vertices}, $\delta g_{i_1\dots i_n}$. The
above defined standard expansion to compute the full Green's
functions corresponds to the Feynman rules associated to
the theory ``$S_Q+(S_I+\delta S)$'', i.e., with $g_{i_1\cdots i_n}+\delta
g_{i_1\cdots i_n}$ as new vertices. Equivalently, one can use an
obvious generalization of the Feynman rules, using one kind of
line, $s^{ij}$, and two kinds of vertices, $g_{i_1\dots i_n}$ and
$\delta g_{i_1\dots i_n}$, which should be considered as
distinguishable. As an alternative, we seek instead a diagrammatic
calculation in terms of $\Gamma[\phi]$ and $\delta S[\phi]$, that is,
using $\overline{s}^{ij}$ as line and $\overline{g}_{i_1\dots i_n}$
and $\delta g_{i_1\dots i_n}$ as vertices. The question of which new
Feynman rules are to be used with these building blocks is answered by
the following

{\bf Theorem.} {\em The Green's functions associated to
$S[\phi]+\delta S[\phi]$ follow from applying the Feynman rules of the
theory ``$\Gamma_Q+(\Gamma_I+\delta S)$'' plus the further
prescription of removing the graphs that contain ``unperturbed
loops'', i.e., loops constructed entirely from effective elements
without any perturbative vertex $\delta g_{i_1\dots i_n}$.}

This constitutes the basic result of this paper. The same statement
holds in the presence of fermions. The proof is given below. We remark
that the previous result does not depend on particular choices, such
as $\overline{g}_i=\overline{g}_{ij}=0$. As a consistency check of the
rules, we note that when $\delta S$ vanishes only tree level graphs of
the theory ``$\Gamma_Q+\Gamma_I$'' remain, which is indeed the correct
result. On the other hand, when $S[\phi]$ is quadratic, it coincides
with its effective action (up to an irrelevant constant) and therefore
there are no unperturbed loops to begin with. Thus, in this case our
rules reduce to the ordinary ones. In this sense, the new rules given
here are the general ones whereas the usual rules correspond only to
the particular case of perturbing an action that is quadratic.

\subsection{Illustration of the new Feynman rules}
To illustrate our rules, let us compute the corrections to the
effective current and the self-energy, $\delta\overline{g}_i$ and
$\delta\Sigma_{ij}$, induced by a perturbation at most quadratic in
the fields, that is,
\begin{equation}
\delta S[\phi]= \delta g_i\phi^i+\frac{1}{2}\delta g_{ij}\phi^i\phi^j \,,
\end{equation}
and at first order in the perturbation. To simplify the result, we
will choose a vanishing $\overline{g}_{ij}$. On the other hand,
$S_Q[\phi]$ will be kept fixed and $\delta S[\phi]$ will be included
in the interacting part of the action, so $\delta\Sigma_{ij}=
\delta\overline{m}_{ij}$.

\begin{figure}[htb]
\centerline{\mbox{\epsfysize=4.6cm\epsffile{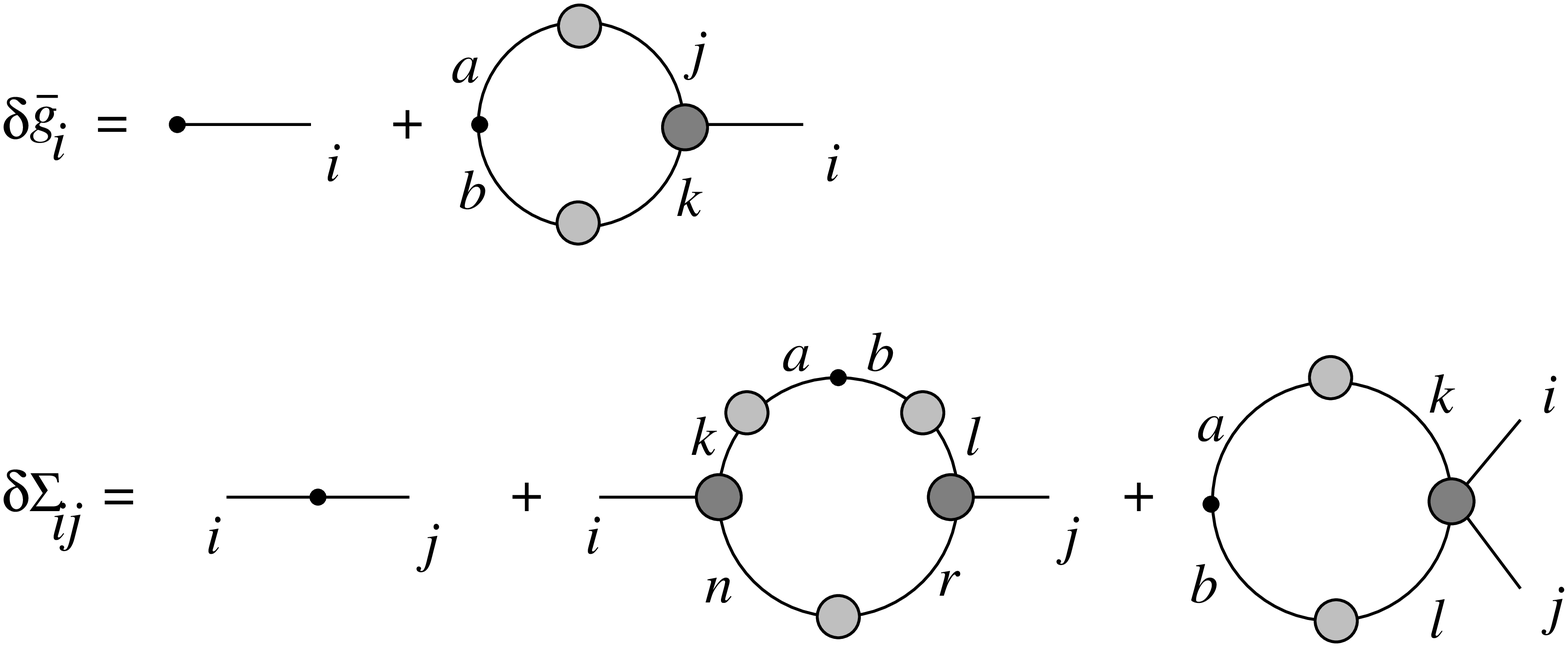}}}
\vspace{6pt}
\caption{Diagrammatic representation of eqs.~(\ref{eq:2}). The small
full dot represents perturbation vertices. All graphs are amputated.}
\label{f:6}
\end{figure}

Applying our rules, it follows that $\delta\overline{g}_i$ is given by
the sum of 1-point diagrams of the theory ``$\Gamma_Q+(\Gamma_I+\delta
S)$'' with either one $\delta g_i$ or one $\delta g_{ij}$ vertex and
which are connected, amputated, 1-particle irreducible and contain no
unperturbed loops. Likewise, $\delta\Sigma_{ij}$ is given by 2-point
such diagrams. It is immediate that $\delta g_i$ can only appear in
0-loop graphs and $\delta g_{ij}$ can only appear in 0- or 1-loop
graphs, since further loops would necessarily be unperturbed. The
following result is thus found
\begin{eqnarray}
\delta\overline{g}_i &=& \delta g_i + \frac{1}{2}\delta g_{ab}
\overline{s}^{aj}\overline{s}^{bk}\overline{g}_{jki}\,, \nonumber \\
\delta\Sigma_{ij} &=& \delta g_{ij} + \delta g_{ab}\overline{s}^{ak}
\overline{s}^{b\ell}\overline{g}_{kni} \overline{g}_{\ell rj}
\overline{s}^{nr} +\frac{1}{2}\delta g_{ab}
\overline{s}^{ak}\overline{s}^{b\ell}\overline{g}_{k\ell ij}\,.
\label{eq:2}
\end{eqnarray}
The graphs corresponding to the r.h.s. are shown in
figure~\ref{f:6}. There, the small full dots represent the
perturbative vertices, the lines with lighter blobs represent the
effective line and the vertices with darker blobs are the effective
vertices. The meaning of this equation is, as usual, that upon
expansion of the skeleton graphs in the r.h.s., every ordinary Feynman
graph (i.e. those of the theory ``$S_Q+(S_I+\delta S)$'') appears once
and only once, and with the correct weight. In other words, the new
graphs are a resummation of the old ones.

\begin{figure}[htb]
\centerline{\mbox{\epsfysize=2.5cm\epsffile{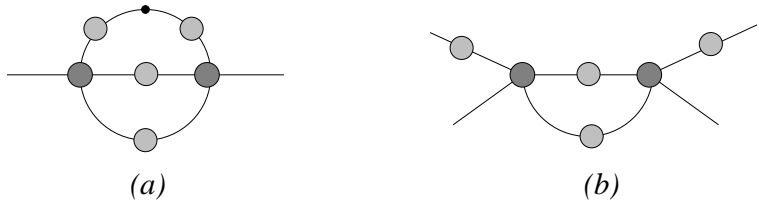}}}
\vspace{6pt}
\caption{$(a)$ A redundant graph. Meaning of lines and vertices as
in figures~\ref{f:5} and ~\ref{f:6}. $(b)$ The associated
unperturbed graph to $(a)$.}
\label{f:7}
\end{figure}

Let us take advantage of the above example to make several
remarks. First, in order to use our rules, all $n$-point effective
vertices have to be considered, in principle. In the example of
figure~\ref{f:6}, only the 3-point proper vertex is needed for the
first order perturbation of the effective current and only the 3- and
4-point proper vertices are needed for the self-energy. Second, after
the choice $\overline{g}_{ij}=0$, the corrections to any proper vertex
requires only a finite number of diagrams, for any given order in each
of the perturbation vertices $\delta g_{i_1\dots i_n}$. Finally,
skeleton graphs with unperturbed loops should not be
included. Consider, e.g. the graph in figure~\ref{f:7}$a$. This graph
contains an unperturbed loop. If its unperturbed loop is contracted to
a single blob, this graph becomes the third 2-point graph in
figure~\ref{f:6}, therefore it is intuitively clear that it is
redundant. In fact, the ordinary graphs obtained by expanding the
blobs in figure~\ref{f:7}$a$ in terms of ``$S_Q+S_I$'' are already be
accounted for by the expansion of the third 2-point graph in
figure~\ref{f:6}.

For a complicated diagram of the theory ``$\Gamma_Q+(\Gamma_I+\delta
S)$'', the cleanest way to check for unperturbed loops is to construct
its {\em associated unperturbed graph}. This is the graph of the theory
``$\Gamma_Q+\Gamma_I$'' which is obtained after deleting all perturbation
vertices, so that the ends previously attached to such vertices become
external legs in the new graph. Algebraically this means to remove the
$\delta g_{i_1\dots i_n}$ factors so that the involved indices become
external (uncontracted) indices. The number of unperturbed loops of
the old (perturbed) graph coincides the number of loops of the
associated unperturbed graph. The associated graph to that in
figure~\ref{f:7}$a$ is depicted in figure~\ref{f:7}$b$.

\section{Some applications}
Of course, the success of the standard Feynman diagrammatic technique
is based on the fact that quadratic actions, unlike non quadratic
ones, can be easily and fully solved. Nevertheless, even when the
theory $S[\phi]$ is not fully solved, our expansion can be
useful. First, it helps in organizing the calculation. Indeed, in the
standard expansion the same 1-, 2-,..., $n$-point unperturbed Green's
functions are computed over and over, as subgraphs, instead of only
once. Second, and related, because the perturbative expansion in
$S_I[\phi]$ must be truncated, in the standard expansion one is in
general using different approximations for the same Green's functions
of $S[\phi]$ in different subgraphs. As a consequence, some known
exact properties (such as symmetries, experimental values of masses or
coupling constants, etc) of the Green's functions of $S[\phi]$ can be
violated by the standard calculation.  On the contrary, in the
expansion proposed here, the Green's functions of $S[\phi]$ are taken
as an input and hence one can make approximations to them (not
necessarily perturbative) to enforce their known exact properties. As
an example consider the Casimir effect. The physical effect of the
conductors is to change the photon boundary conditions. This in turn
corresponds to modify the free photon propagator~\cite{Bo85}, i.e., to
add a quadratic perturbation to the Lagrangian of quantum
electrodynamics (QED). Therefore our expansion applies. The advantage
of using it is that one can first write down rigorous relations
(perturbative in $\delta S$ but non perturbative from the point of
view of QED) and, in a second step, the required QED propagators and
vertex functions can be approximated (either perturbatively or by some
other approach) in a way that is consistent with the experimentally
known mass, charge and magnetic moment of the electron, for instance.
Another example would be chiral perturbation theory: given some
approximation to massless Quantum Chromodynamics (QCD), the
corrections induced by the finite current quark masses can be
incorporated using our scheme as a quadratic perturbation. Other
examples would be the corrections induced by a non vanishing
temperature or density, both modifying the propagator.

\subsection{Derivation of diagrammatic identities}
\begin{figure}[htb]
\centerline{\mbox{\epsfysize=4.3cm\epsffile{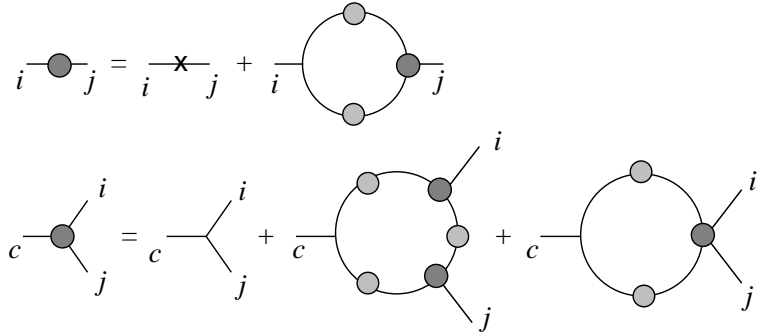}}}
\vspace{6pt}
\caption{Two Schwinger-Dyson equations for a cubic action.}
\label{f:8}
\end{figure}

Another type of applications comes in the derivation of diagrammatic
identities.  We can illustrate this point with some Schwinger-Dyson
equations~\cite{Dy49,Sc51,It80}. Let $\epsilon^i$ be field
independent. Then, noting that the action $S[\phi+\epsilon]$ has
$\Gamma[\phi+\epsilon]$ as its effective action, and for infinitesimal
$\epsilon^i$, it follows that the perturbation $\delta
S[\phi]=\epsilon^i\partial_i S[\phi]$ yields a corresponding
correction $\delta\Gamma[\phi]=\epsilon^i\partial_i \Gamma[\phi]$ in
the effective action. Therefore for this variation we can write:
\begin{eqnarray}
\delta\overline{g}_i &=& \delta\partial_i\Gamma[0]=
\epsilon^j\partial_i\partial_j\Gamma[0]=
\epsilon^j(m+\Sigma)_{ij}\,, \nonumber \\
\delta\Sigma_{ij} &=& \delta\partial_i\partial_j\Gamma[0]=
\epsilon^k\partial_i\partial_j\partial_k\Gamma[0]=
\epsilon^k\overline{g}_{ijk}\,. 
\end{eqnarray}
Let us particularize to a theory with a 3-point bare vertex, then
$\delta S[\phi]$ is at most a quadratic perturbation with vertices
$\delta g_j =\epsilon^i(m_{ij}+g_{ij})$ and $\delta g_{jk}=\epsilon^i
g_{ijk}$. Now we can immediately apply eqs.~(\ref{eq:2}) to obtain
the well known Schwinger-Dyson equations
\begin{eqnarray}
\Sigma_{ij} &=& g_{ij}+
\frac{1}{2}g_{iab}\bar{s}^{a\ell}\bar{s}^{br}\bar{g}_{\ell rj} \,, \\
\overline{g}_{cij} &=& g_{cij}+
g_{cab}\overline{s}^{ak}\overline{s}^{b\ell}\overline{g}_{kni}
\overline{g}_{\ell rj}\overline{s}^{nr}
+\frac{1}{2}g_{cab}\overline{s}^{ak}\overline{s}^{b\ell}
\overline{g}_{k\ell ij}\,. \nonumber
\end{eqnarray}
The corresponding diagrams are depicted in figure~\ref{f:8}.

\subsection{Effective Lagrangians and the double-counting problem}
There are instances in which we do not have (or is not practical to
use) the underlying unperturbed action and we are provided directly,
through the experiment, with the Green's functions. In these cases it
is necessary to know which Feynman rules to use with the exact Green's
functions of $S$. Consider for instance the propagation of particles
in nuclear matter. This is usually described by means of so called
effective Lagrangians involving the nucleon field and other relevant
degrees of freedom (mesons, resonances, photons, etc). These
Lagrangians are adjusted to reproduce at tree level the experimental
masses and coupling constants. (Of course, they have to be
supplemented with form factors for the vertices, widths for the
resonances, etc, to give a realistic description, see
e.g.~\cite{Er88}.) Thus they are a phenomenological approximation to
the effective action rather than to the underlying bare action $S$. So
to say, Nature has solved the unperturbed theory (in this case the
vacuum theory) for us and one can make experimental statements on the
exact (non perturbative) Green's functions. The effect of the nuclear
medium is accounted for by means of a Pauli blocking correction to the
nucleon propagator in the vacuum, namely,
\begin{eqnarray}
G(p)&=&(p^0-\epsilon(\vec{p})+i\eta)^{-1}+
2i\pi n(\vec{p})\delta(p^0-\epsilon(\vec{p})) 
= G_0(p) +\delta G(p)\,,
\end{eqnarray}
where $G_0(p)$ and $G(p)$ stand for the nucleon propagator at vacuum
and at finite density, respectively, $n(\vec{p})$ is the Fermi sea
occupation number and $\epsilon(\vec{p})$ is the nucleon kinetic
energy. In the present case, the vacuum theory is the unperturbed one
whereas the Pauli blocking correction is a 2-point perturbation to the
action and our expansion takes the form of a density expansion.

\begin{figure}[htb]
\centerline{\mbox{\epsfysize=3.0cm\epsffile{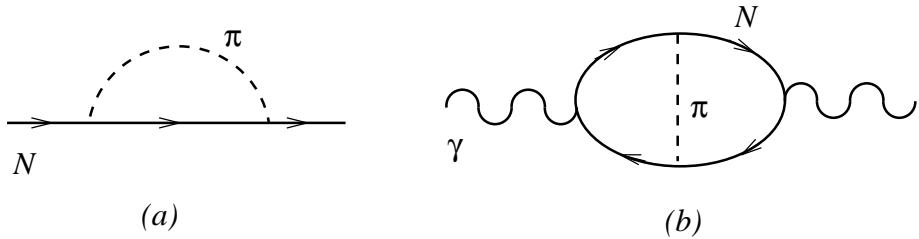}}}
\vspace{6pt}
\caption{Nucleon (a) and photon (b) self-energy diagrams.}
\label{f:9}
\end{figure}

The use of an effective Lagrangian, instead of a more fundamental one,
allows to perform calculations in terms of physical quantities and
this makes the phenomenological interpretation more direct. However,
the use of the standard Feynman rules is not really justified since
they apply to the action and not to the effective action, to which the
effective Lagrangian is an approximation. A manifestation of this
problem comes in the form of double-counting of vacuum contributions,
which has to be carefully avoided. This is obvious already in the
simplest cases. Consider, for instance, the nucleon self-energy coming
from exchange of virtual pions, with corresponding Feynman graph
depicted in figure~\ref{f:9}$a$. This graph gives a non vanishing
contribution even at zero density. Such vacuum contribution is
spurious since it is already accounted for in the physical mass of the
nucleon. The standard procedure in this simple case is to subtract the
same graph at zero density in order to keep the true self-energy. This
is equivalent to drop $G_0(p)$ in the internal nucleon propagator and
keep only the Pauli blocking correction $\delta G(p)$. In more
complicated cases simple overall subtraction does not suffice, as it
is well known from renormalization theory; there can be similar
spurious contributions in subgraphs even if the graph vanishes at zero
density. An example is shown in the photon self-energy graph of
figure~\ref{f:9}$b$. The vertex correction subgraphs contain a purely
vacuum contribution that is already accounted for in the effective
$\gamma NN$ vertex. Although such contributions vanish if the
exchanged pion is static, they do not in general. As is clear from our
theorem, the spurious contributions are avoided by not allowing vacuum
loops in the graphs. That is, for each standard graph consider all the
graphs obtained by substituting each $G(p)$ by either $G_0(p)$ or
$\delta G(p)$ and drop all graphs with any purely vacuum loop. We
emphasize that strictly speaking the full propagator and the full
proper vertices of the vacuum theory have to be used to construct the
diagrams. In each particular application it is to be decided whether a
certain effective Lagrangian (plus form factors, widths, etc) is a
sufficiently good approximation to the effective action.

\subsection{Derivation of low density theorems}
A related application of our rules comes from deriving low density
theorems. For instance, consider the propagation of pions in nuclear
matter and in particular the pionic self-energy at lowest order in an
expansion on the nuclear density. To this end one can use the first
order correction to the self-energy as given in eq.~(\ref{eq:2}), when
the labels $i,j$ refer to pions and the 2-point perturbation is the
Pauli blocking correction for the nucleons. Thus, the labels
$a,b,k,\ell$ (cf. second line of figure~\ref{f:6}) necessarily refer
to nucleons whereas $n,r$ can be arbitrary baryons ($B$). In this
case, the first 2-point diagram in figure~\ref{f:6} vanishes since
$i,j$ are pionic labels which do not have Pauli blocking. On the other
hand, as the nuclear density goes to zero, higher order diagrams
(i.e. with more than one full dot, not present in figure~\ref{f:6})
are suppressed and the second and third 2-point diagrams are the
leading contributions to the pion self energy. The $\pi NB$ and
$\pi\pi NN$ proper vertices in these two graphs combine to yield the
$\pi N$ $T$-matrix, as is clear by cutting the corresponding graphs by
the full dots. (Note that the Dirac delta in the Pauli blocking term
places the nucleons on mass shell.) We thus arrive at the following
low density theorem~\cite{Hu75}: at lowest order in a density
expansion in nuclear matter, the pion optical potential is given by
the nuclear density times the $\pi N$ forward scattering
amplitude. This result holds independently of the detailed
pion-nucleon interaction and regardless of the existence of other kind
of particles as well since they are accounted for by the $T$-matrix.

\subsection{Applications to non perturbative renormalization in
Quantum Field Theory}
Let us consider a further application, this time
to the problem of renormalization in Quantum Field Theory (QFT). To be
specific we consider the problem of ultraviolet divergences. To first
order in $\delta S$, our rules can be written as
\begin{equation}
\delta\Gamma[\phi] =\langle\delta S\rangle^\phi\,,
\label{eq:Lie}
\end{equation}
where $\langle A\rangle^\phi$ means the expectation value of $A[\phi]$
in the presence of an external current $J$ tuned to yield $\phi$ as
the expectation value of the field. This formula is most simply
derived directly from the definitions give above. (In passing, let us
note that this formula defines a group of transformations in the space
of actions, i.e., unlike standard perturbation theory, it preserves
its form at any point in that space.) We can consider a family of
actions, taking the generalized coupling constants as parameters, and
integrate the above first order evolution equation taking e.g. a
quadratic action as starting point. Perturbation theory corresponds to
a Taylor expansion solution of this equation.

To use this idea in QFT, note that our rules directly apply to any
pair of regularized bare actions $S$ and $S+\delta S$. Bare means that
$S$ and $S+\delta S$ are the true actions that yield the expectation
values in the most naive sense and regularized means that the cut off
is in place so that everything is finite and well defined. As it is
well known, a parametric family of actions is said to be
renormalizable if the parameters can be given a suitable dependence on
the cut off so that all expectation values remain finite in the limit
of large cut off (and the final action is non trivial, i.e., non
quadratic). In this case the effective action has also a finite
limit. Since there is no reason to use the same cut off for $S$ and
$\delta S$, we can effectively take the infinite cut off limit in
$\Gamma$ keeping finite that of $\delta S$. (For instance, we can
regularize the actions by introducing some non locality in the
vertices and taking the local limit at different rates for both
actions.) So when using eq.~(\ref{eq:Lie}), we will find diagrams with
renormalized effective lines and vertices from $\Gamma$ and bare
regularized vertices from $\delta S$. Because $\delta\Gamma$ is also
finite as the cut off is removed, it follows that the divergences
introduced by $\delta S$ should cancel with those introduced by the
loops. This allows to restate the renormalizability of a family of
actions as the problem of showing that 1) assuming a given asymptotic
behaviour for $\Gamma$ at large momenta, the parameters in $\delta S$
can be given a suitable dependence on the cut off so that
$\delta\Gamma$ remains finite, 2) the assumed asymptotic behaviour is
consistent with the initial condition (e.g. a free theory) and 3) this
asymptotic behaviour is preserved by the evolution equation. This
would be an alternative to the usual forest formula analysis which
would not depend on perturbation theory.  If the above program were
successfully carried out (the guessing of the correct asymptotic
behaviour being the most difficult part) it would allow to write a
renormalized version of the evolution equation~(\ref{eq:Lie}) and no
further renormalizations would be needed. (Related ideas regarding
evolution equations exist in the context of low momenta expansion, see
e.g.~\cite{Morris} or to study finite temperature
QFT~\cite{Pietroni}.)

\begin{figure}[htb]
\centerline{\mbox{\epsfxsize=12cm\epsffile{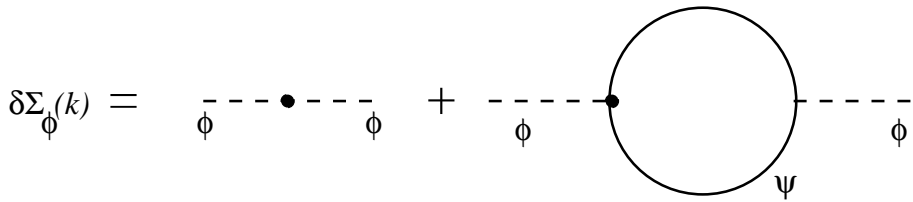}}}
\vspace{6pt}
\caption{Diagrammatic representation of eq.~(\ref{eq:21}).}
\label{f:10}
\end{figure}

To give an (extremely simplified) illustration of these ideas, let us
consider the family of theories with Euclidean action
\begin{equation}
S[\phi,\psi]=\int
d^4x\left(\frac{1}{2}(\partial\phi)^2+\frac{1}{2}m^2\phi^2
+\frac{1}{2}(\partial\psi)^2+\frac{1}{2}M^2\psi^2
+\frac{1}{2}g\phi\psi^2 + h\phi + c\right).
\end{equation}
Here $\phi(x)$ and $\psi(x)$ are bosonic fields in four dimensions.
Further, we will consider only the approximation of no
$\phi$-propagators inside of loops. This approximation, which treats
the field $\phi$ at a quasi-classical level, is often made in the
literature. It As it turns out, the corresponding evolution equation
is consistent, that is, the right-hand side of eq.~(\ref{eq:Lie}) is
still an exact differential after truncation.  In order to evolve the
theory we will consider variations in $g$, and also in $c$, $h$ and
$m^2$, since these latter parameters require a ($g$-dependent)
renormalization. (There are no field, $\psi$-mass or coupling constant
renormalization in this approximation.) That is
\begin{equation}
\delta S[\phi,\psi]= \int
d^4x\left(\frac{1}{2}\delta m^2\phi^2
+\frac{1}{2}\delta g\phi\psi^2 + \delta h\phi + \delta c\right)\,.
\end{equation}
The graphs with zero and one $\phi$-leg are divergent and clearly they
are renormalized by $\delta c$ and $\delta h$, so we concentrate on the
remaining divergent graph, namely, that with two $\phi$-legs. Noting
that in this quasi-classical approximation $g$ coincides with the full
effective coupling constant and $S_\psi(q)=(q^2+M^2)^{-1}$ coincides
with the the full propagator of $\psi$, an application of the rules
gives (cf. figure~\ref{f:10})
\begin{equation}
\delta\Sigma_\phi(k)= \delta m^2 - \delta g
g\int\frac{d^4q}{(2\pi)^4}\theta(\Lambda^2-q^2) S_\psi(q)S_\psi(k-q)\,,
\label{eq:21}
\end{equation}
where $\Lambda$ is a sharp ultraviolet cut off.
Let us denote the cut off integral by $I(k^2,\Lambda^2)$. This
integral diverges as $\frac{1}{(4\pi)^2}\log(\Lambda^2)$ for large
$\Lambda$ and fixed $k$. Hence $\delta\Sigma_\phi$ is guaranteed to
remain finite if, for large $\Lambda$, $\delta m^2$ is taken in the
form
\begin{equation}
\delta m^2 = \delta m_R^2 + \delta g g\frac{1}{(4\pi)^2}\log(\Lambda^2/\mu^2)
\end{equation}
where $\mu$ is an arbitrary scale (cut off independent), and $\delta
m_R^2$ is an arbitrary variation. Thus, the evolution equation for
large cut off can be written in finite form, that is, as a
renormalized evolution equation, as follows
\begin{equation}
\delta\Sigma_\phi(k)= \delta m_R^2 - \delta g gI_R(k^2,\mu^2)\,,
\end{equation}
where
\begin{equation}
I_R(k^2,\mu^2)=\lim_{\Lambda\to\infty}\left(I(k^2,\Lambda^2)
-\frac{1}{(4\pi)^2}\log(\Lambda^2/\mu^2)\right)\,.
\end{equation}
Here $\delta g$ and $\delta m_R^2$ are independent and arbitrary
ultraviolet finite variations. The physics remains constant if a
different choice of $\mu$ is compensated by a corresponding change in
$\delta m_R^2$ so that $\delta m^2$, and hence the bare regularized action,
is unchanged. The essential point has been that $\delta m^2$ could be
chosen $\Lambda$ dependent but $k^2$ independent. As mentioned, this
example is too simple since it hardly differs from standard
perturbation theory. The study of the general case (beyond
quasi-classical approximations) with this or other actions seems very
interesting from the point of view of renormalization theory.

\section{Proof of the theorem}
In order to prove the theorem it will be convenient to change the
notation: we will denote the unperturbed action by $S_0[\phi]$ and its
effective action by $\Gamma_0[\phi]$. The generating function of the
full perturbed system is
\begin{equation}
Z[J]= \int\exp\left(S_0[\phi]+\delta S[\phi] +J\phi\right)\,d\phi \,.
\label{eq:8}
\end{equation}
By definition of the effective
action, the connected generating function of the unperturbed theory is
\begin{equation}
W_0[J]=\max_\phi\left(\Gamma_0[\phi]+J\phi\right)\,,
\end{equation}
thus, up to a constant ($J$-independent) factor, we can write
\begin{eqnarray}
\exp\left(W_0[J]\right) &=& \lim_{\hbar\to 0}\left[\int
\exp\left(\hbar^{-1}\left(\Gamma_0[\phi]+J\phi\right)\right)
\,d\phi\right]^{\textstyle\hbar}\,.
\end{eqnarray}
$\hbar$ is merely a bookkeeping parameter here which is often used to
organize the loop expansion~\cite{Co73,Ne87}. The $\hbar$-th power above can be
produced by means of the replica method~\cite{Ne87}. To this end we
introduce a number $\hbar$ of replicas of the original field, which
will be distinguished by a new label $k$. Thus, the previous equation
can be rewritten as
\begin{equation}
\exp\left(W_0[J]\right)= \lim_{\hbar\to 0}\int
\exp\left(\hbar^{-1}\sum_k\left(\Gamma_0[\phi_k]+J\phi_k\right)\right)
\prod_kd\phi_k \,.
\label{eq:10}
\end{equation}
On the other hand, the identity (up to a constant)
$\int\exp\left(J\phi\right)\,d\phi = \delta[J]$, where $\delta[J]$
stands for a Dirac delta, allows to write the reciprocal relation of
eq.~(\ref{eq:Z}), namely
\begin{equation}
\exp\left(S_0[\phi]\right)= \int\exp\left(W_0[J_0]-J_0\phi\right)\,dJ_0 \,.
\label{eq:11}
\end{equation}
If we now use eq.~(\ref{eq:10}) for $\exp W_0$ in eq.~(\ref{eq:11})
and the result is substituted in eq.~(\ref{eq:8}), we obtain
\begin{equation}
Z[J]= \lim_{\hbar\to 0}\int\exp\left(
\hbar^{-1}\sum_k\left(\Gamma_0[\phi_k]+J_0\phi_k\right) +\delta
S[\phi] +\left(J-J_0\right)\phi \right) \,dJ_0\,d\phi\prod_kd\phi_k
\,.
\end{equation}
The integration over $J_0$ is immediate and yields a Dirac delta for
the variable $\phi$, which allows to carry out also this
integration. Finally the following formula is obtained:
\begin{equation}
Z[J]= \lim_{\hbar\to 0}\int
\exp\left(\hbar^{-1}\sum_k\left(\Gamma_0[\phi_k]+J\phi_k\right)
+\delta S\big[\hbar^{-1}\sum_k\phi_k\big]\right)\prod_kd\phi_k \,.
\label{eq:13}
\end{equation}
which expresses $Z[J]$ in terms of $\Gamma_0$ and $\delta S$. Except
for the presence of replicas and explicit $\hbar$ factors, this
formula has the same form as that in eq.~(\ref{eq:8}) and hence it
yields the same standard Feynman rules but with effective lines and
vertices.

Consider any diagram of the theory ``$\Gamma_Q+(\Gamma_I+\delta S)$'',
as described by eq.~(\ref{eq:13}) before taking the limit $\hbar\to
0$. Let us now show that such diagram carries precisely a factor
$\hbar^{L_0}$, where $L_0$ is the number of unperturbed loops in the
graph. Let $P$ be the total number of lines (both internal and
external), $E$ the number of legs, $L$ the number of loops and $C$ the
number of connected components of the graph. Furthermore, let $V^0_n$
and $\delta V_n$ denote the number of $n$-point vertices of the types
$\Gamma_0$ and $\delta S$ respectively. After these definitions, let
us first count the number of $\hbar$ factors coming from the explicit
$\hbar^{-1}$ in eq.~(\ref{eq:13}). The arguments are
standard~\cite{Co73,It80,Ne87}: from the Feynman rules it is clear
that each $\Gamma_0$ vertex carries a factor $\hbar^{-1}$, each
effective propagator carries a factor $\hbar$ (since it is the inverse
of the quadratic part of the action), each $n$-point $\delta S$ vertex
carries a factor $\hbar^{-n}$ and each leg a $\hbar^{-1}$ factor
(since they are associated to the external current $J$). That is, this
number is
\begin{equation}
N_0 = P - \sum_{n\geq 0} V^0_n-E-\sum_{n\ge 0}n\delta V_n \,.
\end{equation}
Recall now the definition given above of the associated unperturbed
diagram, obtained after deleting all perturbation vertices, and let
$P_0$, $E_0$, $L_0$ and $C_0$ denote the corresponding quantities for
such unperturbed graph. Note that the two definitions given for the
quantity $L_0$ coincide. Due to its definition, $P_0=P$ and also
$E_0=E+\sum_{n\geq 0}n\delta V_n$, this allows to rewrite $N_0$ as
\begin{equation}
N_0= P_0-\sum_{n\geq 0} V^0_n-E_0\,.
\end{equation}
Since all quantities now refer a to the unperturbed graph, use can be
made of the well known diagrammatic identity $N_0=L_0-C_0$. Thus from
the explicit $\hbar$, the graph picks up a factor
$\hbar^{L_0-C_0}$. Let us now turn to the implicit $\hbar$ dependence
coming from the number of replicas. The replica method idea applies
here directly: because all the replicas are identical, summation over
each different free replica label in the diagram yields precisely one
$\hbar$ factor. From the Feynman rules corresponding to the theory of
eq.~(\ref{eq:13}) it is clear that all lines connected through
$\Gamma_0$ vertices are constrained to have the same replica label,
whereas the coupling through $\delta S$ vertices does not impose any
conservation law of the replica label. Thus, the number of different
replica labels in the graph coincides with $C_0$. In this argument is
is essential to note that the external current $J_i$ has not been
replicated; it couples equally to all the replicas. Combining this
result with that previously obtained, we find that the total $\hbar$
dependence of a graph goes as $\hbar^{L_0}$. As a consequence, all
graphs with unperturbed loops are removed after taking the limit
$\hbar\to 0$. This establishes the theorem.

Some remarks can be made at this point. First, it may be noted that
some of the manipulations carried out in the derivation of
eq.~(\ref{eq:13}) were merely formal (beginning by the very definition
of the effective action, since there could be more than one extremum
in the Legendre transformation), however they are completely
sufficient at the perturbative level. Indeed, order by order in
perturbation theory, the unperturbed action $S_0[\phi]$ can be
expressed in terms of its effective action $\Gamma_0[\phi]$, hence the
Green's functions of the full theory can be expressed perturbatively
within the diagrams of the theory ``$\Gamma_Q+(\Gamma_I+\delta
S)$''. It only remains to determine the weighting factor of each graph
which by construction (i.e. the order by order inversion) will be just
a rational number.  Second, it is clear that the manipulations that
lead to eq.~(\ref{eq:13}) can be carried out in the presence of
fermions as well, and the same conclusion applies. Third, note that in
passing, it has been proven also the statement that the effective
action yields at tree level the same Green's functions as the bare
action at all orders in the loop expansion, since this merely
corresponds to set $\delta S[\phi]$ to zero. Finally,
eq.~(\ref{eq:13}) does not depend on any particular choice, such as
fixing $\langle\phi^i\rangle=0$ to remove tadpole subgraphs.

\section*{Acknowledgments}
L.L. S. would like to thank C. Garc\'{\i}a-Recio and J.W. Negele for
discussions on the subject of this paper. This work is supported in
part by funds provided by the U.S. Department of Energy (D.O.E.)
under cooperative research agreement \#DF-FC02-94ER40818, Spanish
DGICYT grant no. PB95-1204 and Junta de Andaluc\'{\i}a grant no.
FQM0225.

\end{document}